\documentclass[aps,prb,reprint, superscriptaddress,showpacs]{revtex4-1}
\usepackage{amsmath}
\usepackage{amssymb}
\usepackage{graphicx}

\begin{document}

\title{Nonreciprocal transmission of neutrons through the noncoplanar magnetic system}
\author{D.A. Tatarskiy}
\email[tatarsky@ipmras.ru]{}
\affiliation{Institute for Physics of Microstructures RAS, GSP-105, Nizhny Novgorod, 603950, Russia}
\affiliation{Lobachevsky State University of Nizhny Novgorod, Gagarin 23, Nizhniy Novgorod, 603950, Russia}
\author{A.V. Petrenko}
\affiliation{Joint Institute for Nuclear Research, Joliot-Curie St. 6, Dubna, Moscow region, 141980, Russia}
\author{S.N. Vdovichev}
\affiliation{Institute for Physics of Microstructures RAS, GSP-105, Nizhny Novgorod, 603950, Russia}
\affiliation{Lobachevsky State University of Nizhny Novgorod, Gagarin 23, Nizhniy Novgorod, 603950, Russia}
\author{O.G. Udalov}
\affiliation{Institute for Physics of Microstructures RAS, GSP-105, Nizhny Novgorod, 603950, Russia}
\affiliation{Department of Physics and Astronomy, California State University Northridge, Northridge, California 91330, USA}
\author{Yu.V. Nikitenko}
\affiliation{Joint Institute for Nuclear Research, Joliot-Curie St. 6, Dubna, Moscow region, 141980, Russia}
\author{A.A. Fraerman}
\affiliation{Institute for Physics of Microstructures RAS, GSP-105, Nizhny Novgorod, 603950, Russia}
\affiliation{Lobachevsky State University of Nizhny Novgorod, Gagarin 23, Nizhniy Novgorod, 603950, Russia}

\date{\today}

\begin{abstract}
We report on observation of the time reversal symmetry breaking in an unpolarized neutrons scattering experiment based on a nonreciprocal cell consisting of two magnetic mirrors placed in an external magnetic field. The neutron transmittivity through the system is measured. The time reversal symmetry holds for coplanar magnetic configuration, meaning that the transmission does not vary when the neutron source and detector are interchanged. In the case of a noncoplanar magnetic structure, though, the time reversal symmetry breaks down. Understanding the origin of the nonreciprocity of scattering for spin 1/2 particles opens up new possibilities in the field of nonreciprocal spintronic device development.
\end{abstract}


\maketitle

\section{Introduction}Behavior of a spin 1/2 particle in an inhomogeneous magnetic or exchange field is one of the most fundamental problems in condensed matter physics. The Schr\"{o}dinger equation for an electron interacting with the exchange field of ferromagnet provides the basis for spintronics which is an actively growing area of fundamental and applied research currently~\cite{Pulizzi2012}. Numerous physical phenomena in magnetic systems are described using this equation: giant magneto-resistance~\cite{Fert,Grunberg}, topological Hall effect~\cite{AharonovStern,Skyrmions}, current induced torques~\cite{Berger,Slonczewski,Zhang} and etc. A similar equation describes neutron behavior in magnetic medium~\cite{Izyumov} which is the subject topic of magnetic neutronography. Polarization~\cite{Izyumov} and Zeeman splitting of a neutron beam in noncollinear field~\cite{Ignatovich,Felcher,Korneev}, neutron spin echo~\cite{Mezei} and etc. can be understood using Schr\"{o}dinger equation with Pauli term

\begin{figure}
\includegraphics[width=0.7\linewidth]{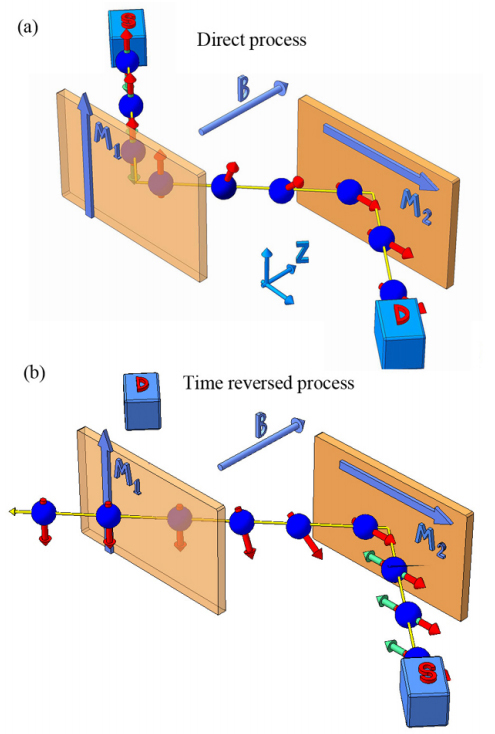}
\caption{\label{main}(Color online) (a,b) Demonstration experiment of nonreciprocal neutron transmission. Unpolarized neutrons are emitted from a source (S) and registered with a detector (D). The beam successively reflects from mirrors with magnetizations $\mathbf M_1$ and $\mathbf M_2$. The neutron spin precesses in external magnetic field $\mathbf B$ between the mirrors. When magnetizations are perpendicular the successively reflection of the neutron beam are different in the direct (a) and time-reversed (b) processes.}
\end{figure}

\begin{equation}
i\hbar\frac{\partial\hat{\psi}}{\partial t}=\left[ \left( \frac{\hat{\mathbf{p}}^2}{2 m} + V \left( \mathbf{r} \right) \right) \hat{I} -\mu _n \left( \hat{\mathbf{\sigma}} \cdot \mathbf{B} \left( \mathbf{r} \right) \right) \right] \hat{\psi},
\label{schrod}
\end{equation}
where $\hat{\mathbf{p}}$ is the momentum operator, $V(\mathbf{r})$ is the scalar potential, $\mathbf{B}$ is the magnetic field, $m$ and $\mu_n$ are the mass and the magnetic moment of neutron, $\hat{I}$ is the unit matrix, $\hat{\mathbf{\sigma}}$ is the vector of Pauli matrices and $\hat{\psi}$ is the spinor wave function. The equation does not include interaction of magnetic field with the particle charge. This is valid for neutrons since they are neutral particles. Sometimes it is a reasonable approximation for electrons in ferromagnets where the exchange field is rather strong and one can neglect the Lorentz force. For electrons, one should substitute the exchange constant for magnetic moment $\mu_n$ and unit vector along the local magnetization for magnetic induction.

Let us denote $S(\mathbf{k}',\mathbf{k},\mathbf{B}(\mathbf{r}))$ as a differential cross-section (DCS), i.e. scattering in an element of the solid angle. In terms of scattering of unpolarized particles the time reversal symmetry can be expressed as the reciprocity theorem~\cite{landau3,Reciprocity}
\begin{equation}
S(\mathbf{k},\mathbf{k}',\mathbf{B}(\mathbf{r}))=S(-\mathbf{k}',-\mathbf{k},-\mathbf{B}(\mathbf{r})),
\label{rectheorem}
\end{equation}
where $\mathbf{k},\mathbf{k}'$ are the momenta of the incident and scattered neutrons and $\mathbf{B}(\mathbf{r})$ is the magnetic induction distribution. We refer to the scattering in a certain system is called reciprocal if the DCS is the same for direct $(\mathbf{k}\rightarrow\mathbf{k}')$ and "time-reversed" $(-\mathbf{k}'\rightarrow-\mathbf{k})$ processes (below we use the term "time-reversed" without quotes)
\begin{equation}
S^{\mathrm R}(\mathbf{k},\mathbf{k}',\mathbf{B}(\mathbf{r}))=S^{\mathrm R}(-\mathbf{k}',-\mathbf{k},\mathbf{B}(\mathbf{r})).
\label{Recip}
\end{equation}
Time reversal symmetry breaking and nonreciprocal scattering occur when the direct and time-reversed scattering processes are not the same
\begin{equation}
S^{\mathrm{NR}}(\mathbf{k},\mathbf{k}',\mathbf{B}(\mathbf{r}) )\neq S^{\mathrm{NR}}(-\mathbf{k}',-\mathbf{k},\mathbf{B}(\mathbf{r})).
\label{NRecip}
\end{equation}

One of the best-known examples of nonreciprocal processes is the light transmission through the Faraday cell. The necessary condition for nonreciprocal light scattering is the absence of spatial and time reversal symmetries (presence of magnetic field or magnetization)~\cite{Pikus1992}. The microscopic mechanism for nonreciprocal light scattering are the Lorentz force and spin-orbit interaction. Neither of these effects is the case for neutrons, hence, they do not appear in Eq.~(\ref{schrod}). It means that the requirements for observation of nonreciprocal phenomena for neutrons and other particles described by Eq.~(\ref{schrod}) are more stringent. Below we discuss an additional symmetry of the DCS as a specific property of the Schr\"{o}dinger equation with Pauli term Eq.~(\ref{schrod}). Scattering of light does not have such additional symmetry. In the paper we show that it is possible to create nonreciprocal cell for neutrons.

Understanding the origin of nonreciprocal scattering for spin 1/2 particles opens new possibilities for development of nonreciprocal spintronic devices. Recently a magnetic diode using the nonreciprocal behavior of electrons was proposed~\cite{Udalov2012}. This device consists of two ferromagnetic leads separated by non-magnetic metal layer placed into external magnetic field. The ferromagnetic leads serve as spin polarizer and spin detector. In the middle layer the electrons spins precess around an external field. This geometry is very much the same as what we used in the experiment discussed in the paper.

The time reversal symmetry breaking in Eq.~(\ref{schrod}) leads to several physical phenomena. The topological Hall effect was predicted~\cite{AharonovStern} and observed for electrons in magnetic Skyrmions~\cite{Skyrmions}. A similar effect was also predicted for neutron diffraction on MnSi crystal~\cite{UdalovAphase}. Violation of the time reversal symmetry may lead not only to Hall-like effects, but also to other kinds of nonreciprocal phenomena, such as diode effect and persistent electrical current in mesoscopic magnetic rings with a noncoplanar magnetization distribution~\cite{Tatara, Udalov2008}. Thus, the time reversal symmetry properties of Eq.~(\ref{schrod}) are significant for a wide variety of physical problems.

Due to the strong symmetry of Eq.~(\ref{schrod}) the scattering of neutrons is always reciprocal in systems with coplanar magnetic field even if a system lacks of spatial inversion symmetry. It can be demonstrated as follows. The DCS is symmetric with respect to rotation $\hat{R}_{\mathbf{n}}^{\theta}$ of magnetic induction $\mathbf B$ by the same angle $\theta$ for each point of a system around arbitrary axis $\mathbf{n}$ \cite{Tatarskiy}
\begin{equation}
S(\mathbf{k},\mathbf{k}',\mathbf{B}(\mathbf{r}))=S(\mathbf{k},\mathbf{k}',\hat{R}_{\mathbf{n}}^{\theta}\mathbf{B}(\mathbf{r})).
\label{rotsymm}
\end{equation}
For coplanar magnetic field (e.g. when only $x$ and $z$ components of the field are non-zero) $-\mathbf{B}=\hat{R}_{y}^{\pi}\mathbf{B}$, and in accordance with~(\ref{rectheorem}) and~(\ref{rotsymm}) we arrive at Eq.~(\ref{Recip}). In noncoplanar magnetic systems, the rotation around any axis cannot compensate the inversion of the magnetic field sign and the elastic scattering becomes nonreciprocal.

\section{Toy model} The theory of nonreciprocal reflection of neutrons was proposed in our previous work~\cite{Tatarskiy}. A successive reflection of unpolarized neutrons from three mirrors with noncoplanar magnetizations was shown to be varying when the neutron source and detector were interchanged. We adopt three mirrors system for the demonstrational experiment.

\begin{figure}
\includegraphics[width=0.8\linewidth]{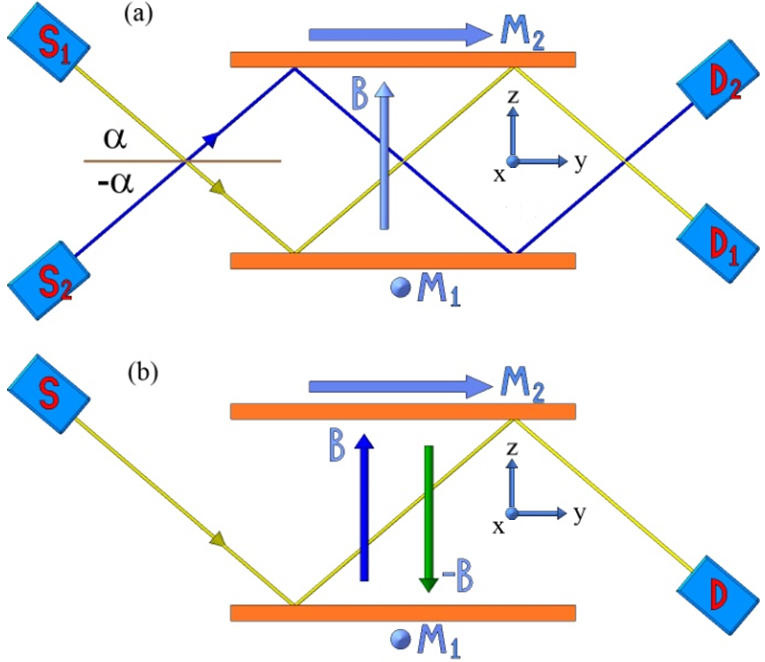}
\caption{\label{nonrecExp} (Color online) Two types of measurements. Two mirrors placed in front of each other (orange slabs). (a) Two measurements with different glancing angles ($\alpha$ and $-\alpha$) but with fixed external magnetic field are performed. Neutron beams from sources $S_1$ and $S_2$ correspond to direct and time-reversed processes (see text). (b) Two measurements with opposite sign of external magnetic field but with fixed source and detector are performed. Beam passage through the system with different sign of magnetic field corresponds to direct and time-reversed processes.}
\end{figure}

\begin{figure}
\includegraphics[width=0.8\linewidth]{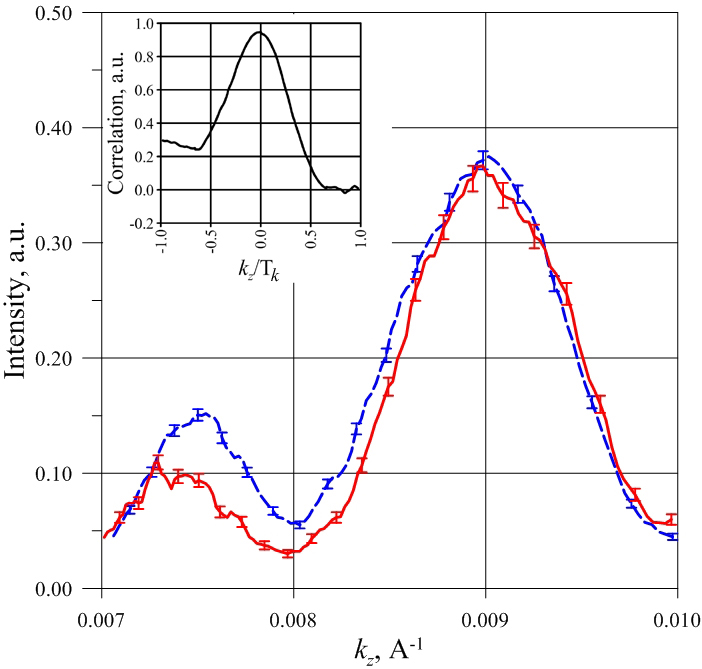}
\caption{\label{C18}(Color online) Neutron transmission through coplanar system in the external field 18 Oe. Red solid line is direct process, blue dashed line is time-reversed one.}
\end{figure}

Consider neutrons successively reflected by two ideal mirrors placed in an external magnetic field (see Fig.~\ref{main}). When the external magnetic field is perpendicular to the mirrors magnetizations the transmission amplitude is expressed as
\begin{equation}
\label{ampl}
\hat{A}=\hat{R}_y\hat{T}_\varphi\hat{R}_x,
\end{equation}
where $\hat{R}_x=\exp\left(i\hat{\sigma}_x \pi/4\right)\left(\hat{I}+\hat{\sigma}_z\right)$ and $\hat{R}_y=\left(\hat{I}+\hat{\sigma}_z\right)\exp\left(-i\hat{\sigma}_y \pi/4\right)$ are the reflection amplitudes from the corresponding mirrors and $\hat{T}_\varphi=\exp\left(i\hat{\sigma}_z \varphi/2\right)$ is the Larmour precession in the external field with the corresponding phase. The amplitude of the time-reversed transmission is obtained by replacing of $\hat{\sigma}_x\leftrightarrow\hat{\sigma}_y$ matrices in~(\ref{ampl}). The intensities of direct and time-reversed transmission are
\begin{equation}
\label{nonrecI}
I_\pm=Tr\left[\hat{\rho}\hat{A}^\dag\hat{A}\right]=\left(1\pm\sin\varphi\right)/4,
\end{equation}where $\hat{\rho}=\hat{I}/2$ is the density matrix of unpolarized neutrons. The transmission in coplanar configuration can be calculated by substituting $\hat{\sigma}_x$ instead of $\hat{\sigma}_y$ in~(\ref{ampl}). The intensities are equal and described by the following expression $I_\pm=\left(1+\cos\varphi\right)/4$.

As we have shown above the cause of nonreciprocity is the noncommutativity of the spin 1/2 algebra. This noncommutativity provides the basis for theoretical prediction of persistent electrical currents in mesoscopic magnetic rings with a noncoplanar magnetization distribution~\cite{Balatsky,Tatara}.

\begin{figure}
\includegraphics[width=0.8\linewidth]{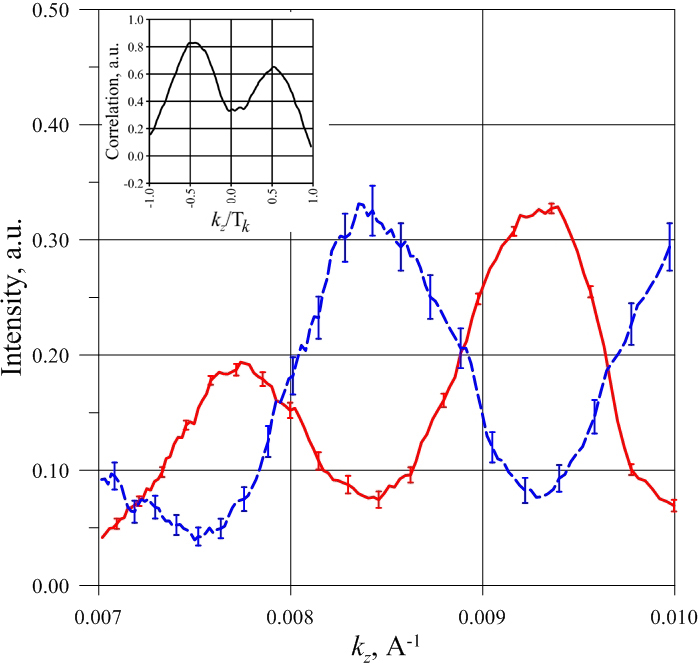}
\caption{\label{NC18} (Color online) Neutron transmission through noncoplanar system in the external field 18 Oe. Red solid line is direct process, blue dashed line is time-reversed one.}
\end{figure}

\section{Experiment}The experimental setup is as follows. Two parallel magnetic mirrors were placed in front of each other (see Fig.~\ref{nonrecExp}). The mirrors were the CoFe films deposited by magnetron sputtering on glass substrates $140\times50$ mm$^2$. The thickness of films is about $115$ nm. The coercive force is about 150 Oe and the remanent magnetization is 90-95\%. Similar magnetic properties of CoFe films were observed earlier~\cite{Underlayer}. High quality glass plate of $d=500$ $\mu$m thickness is placed between the mirrors. The plate controls the intermirror gap and parallelism of the mirrors. The construction is placed in the perpendicular weak homogeneous field induced by two equal magnetic coils. The lateral sizes of the coils are $1000\times200\times150$ mm$^3$ and the external field varies in range 10-30 Oe. The external field is weak, so we can neglect the Zeeman beam-splitting of the reflected neutron beam~\cite{Felcher,Korneev}. Neutrons were emitted from a source S and succesively reflected by the mirrors, as shown in Fig.~\ref{nonrecExp}. Glancing angle $\alpha$ is 7 mrad. This angle is in the interval between the first and second critical angles for 3-6 \AA~neutrons due to the corresponding nuclear and magnetic potentials of CoFe films and substrate (glass). The polarization efficiency of the mirror is about 80\%. Twice reflected neutrons are registered by a detector D. The neutron source for the experiment is pulsed reactor IBR-2M of Joint Institute for Nuclear Research in Dubna.

According to definitions in Eqs.~(\ref{Recip}) and~(\ref{NRecip}) to investigate nonreciprocal effects one needs to measure the intensity of transmitted neutrons before and after the interchange of source and detector. Practically, this means rotation of the entire mirror system as a whole by the angle $\pi$ around the $x$-axis. Such a rotation cannot be performed with a high precision. Therefore we did two types of experiments which are equivalent to measurement with a source-detector interchange.

In the first experiment (see Fig.~\ref{nonrecExp}(a)) we measured the transmittivity of neutron beam incident at the angle $\alpha$. The beam was first reflected by bottom mirror and then by the upper mirror. After that we measured the transmittivity of neutron beam incident at angle $-\alpha$. In this case the order of reflections was different. We can show that this experiment is equal to transmission for direct and time-reversed processes. Imagine a system where the source and detector are interchanged ($S_1\to D_1$ and $D_1\to S_1$). Let us rotate the system as a whole by the angle $\pi$ around the $z$-axis. After such rotation we use the rotational symmetry~(\ref{rotsymm}) and rotate only the magnetic fields by the angle $\pi$ around the $z$-axis. These transformations are equal to the changing of sign of the glancing angle $\alpha$ (Fig.~\ref{nonrecExp}(a)). In the second experiment we measured transmission for fixed source and detector. We did two measurements with different signs of the external magnetic field (see Fig.~\ref{nonrecExp}(b)). Using symmetry relations~(\ref{rotsymm}) and the reciprocity theorem~(\ref{rectheorem}) one can show that this experiment is also equivalent to the measurement of reflectivity for direct and time-reversed processes. We also performed reference experiments with the coplanar system in which magnetizations of both mirrors are along the $y$-axis.

\begin{figure}
\includegraphics[width=0.8\linewidth]{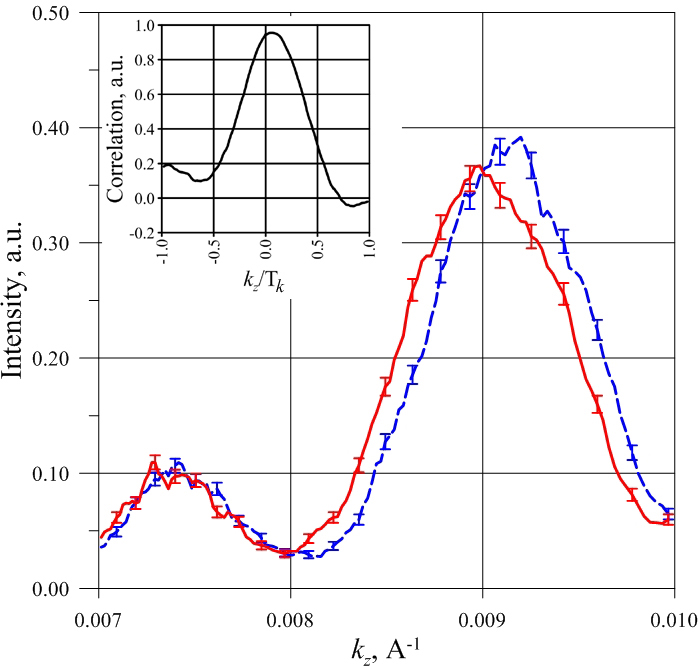}
\caption{\label{C18field}(Color online) Neutron transmission through coplanar system in the external field 18 Oe. Red solid line is direct process, blue dashed line is process with reversed field.}
\end{figure}

\begin{figure}
\includegraphics[width=0.8\linewidth]{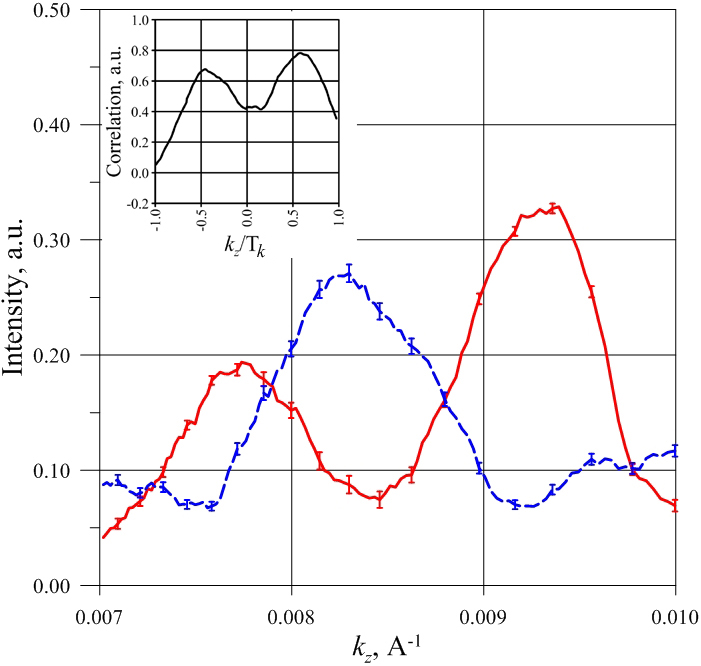}
\caption{\label{NC18field}(Color online) Neutron transmission from noncoplanar system in the external field 18 Oe. Red solid line is direct process, blue dashed line is process with reversed field.}
\end{figure}

Experimental dependencies of the transmission coefficients on the $z$-component of the wave vector are shown in Fig.~\ref{C18}-\ref{NC18field}, where $k_z=\frac{2 \pi}{\lambda}\sin\alpha$ , $\lambda$ is the neutron wavelength. The common property of these curves is the oscillatory behavior. These oscillations originate from the Larmour precession of the magnetic moment of the neutron in the external field. Indeed, the reflection coefficient of the neutron from the second mirror depends on the relative orientation of its magnetic moment and the magnetization direction of the mirror. The reflection is maximal (minimal) when these magnetic moments are parallel (antiparallel). Taking into account, the fact that after reflection from the first mirror the average neutron magnetic moment is perpendicular to the magnetic moment of the second mirror, the condition for observation of maximum transmission through the system has the form $\varphi=\pi \left(n+1/2\right), n=0,1,2...$ where $\varphi=\omega\tau$ is the phase of the magnetic moment of the neutron, precessing with a frequency $\omega=\mu_nB/\hbar\approx 10^5$ 1$/$sec in the external field $B$ during time of flight $\tau=\frac{d\lambda m}{2\pi\hbar\sin\alpha}$ from the first to the second mirror, $d/\sin\alpha=7$ cm is the length of the neutron path between mirrors. In our experiment the neutron magnetic moment makes approximately two full turns between the mirrors
\begin{equation}
\varphi=(\mu_nBdm)/(\hbar^2 k_z)\approx10.
\label{phase}
\end{equation}

Observation of fundamental difference between direct and time-reverse processes in the case of noncoplanar magnetic field distribution is the main result of our work. In Fig.~\ref{C18} and~\ref{NC18} the transmittivity at a different glancing angle sings ($\alpha\rightarrow-\alpha$) for coplanar and noncoplanar configurations are shown. Fig.~\ref{C18field} and~\ref{NC18field} illustrate the transmissions for inverse external field $B\rightarrow-B$. For coplanar magnetic field distribution the direct and time-reversed transmissions are actually identical with our experimental accuracy. The experimental precision is determined by several factor: the goniometer accuracy (0.3 mrad), the neutron beam divergence (about 0.6 mrad) and the external field time fluctuations (0.5 Oe per day). For noncoplanar field distribution the situation is completely different: the maximum transmittivity value in direct process corresponds to the minimal transmittivity in reversed process and vice versa. The relative difference in the transmission coefficients of the direct and inverse processes is up to 75\% (Fig.~\ref{NC18field}). We calculate correlation function to analyze difference between direct (red solid lines) and time-reversed (blue dashed lines) reflections~\cite{hudson}. It is shown on the insets in Fig.~\ref{C18}-\ref{NC18field}. $T_k=\frac{\pi \hbar^2 k^2_z}{\mu_nmd}\frac{1}{B}$ is the quasiperiod of intensity oscillation. The quasiperiod is calculated for $k_z=0.008~\AA^{-1}$. The correlations for the cases of coplanar and noncoplanar magnetic field distribution are different. For coplanar system the correlation has one maximum near zero. In other words the direct and time-reversed transmissions in coplanar field are equal within the accuracy of our experiment. On the contrary the correlations for noncoplanar field have two maxima near one half of phase period. The intensities of direct and time-reversed processes are counter phased, which suggests that the transmission is nonreciprocal.

\section{Discussion} Our experiment is very similar to neutron spin echo (NSE) experiment~\cite{Mezei}. Each magnetic mirror is the boundary with sudden change of the magnetic field because the mirrors' magnetizations are perpendicular to the external (guide) field. In this respect our mirrors are very much the same as  the $\pi/2$-flipper of the NSE experiments as in Ref.~\cite{Mezei}. For all parallel magnetizations we have straightforward measurement of NSE oscillations similar to observed in Ref.~\cite{Mezei}. The conventional NSE experiment means that the $\pi/2$-flippers have parallel magnetic fields. In the noncoplanar configuration the magnetizations of our mirrors are perpendicular to each other and this is the main difference from NSE experiments.

Besides the considered system is a nonreciprocal cell for neutrons similar to the nonreciprocal Faraday cell for light. Both cells consist of a polarizer, analyzer and a phase shifter. In our case they are the magnetic mirrors and magnetic field in the gap between the mirrors. There is a fundamental difference, though, between our system and the Faraday cell, being in that the fact that photons are bosons. The difference in the transmission intensities in the forward and reverse direction for a Faraday cell is proportional to $\sim \sin(2\gamma)$, and for neutrons cell it is proportional to $\sin(\gamma)$, where $\gamma$ is the angle between the polarizer and analyzer (magnetic moments of the mirrors in the considered case). Note that the nonreciprocal cell, similar to the one studied in our work, can be realized for electrons. The spin precession of electron has been observed experimentally~\cite{Jedema} by measuring the conductivity of the conductor between the two ferromagnetic electrodes placed in a perpendicular magnetic field. In these experiments the magnetic moments of the ferromagnetic electrodes were parallel or antiparallel and  the whole magnetic system was coplanar. If the electrodes were arranged perpendicular to each other, the magnetization distribution in the system could be noncoplanar and nonreciprocal effect can be observed. Such a situation was theoretically considered in Ref.~\cite{Udalov2012}. A Rectification effect appearing due to non-commutativity of spin algebra was predicted. The essential peculiarity of electrons is that they propagate through the middle layer diffusively. This leads to relaxation of the electron spin at characteristic length $l_{\mathrm s}$ and restricts the length of the middle layer $d<l_{\mathrm s}$.

\begin{acknowledgments}
Authors are grateful to N.~Korotkova, B.~Gribkov, P.~Yunin and N.~Gusev for assistance in magnetic mirror preparation and attestation. Also authors are grateful to S.~Kozhevnikov, A.~Klimov, V.~Rogov and I.~Shereshevskii for processing the experimental results. Authors thank I.~Beloborodov for useful discussion of the results.

This work was supported by The Ministry of education and science of the Russian Federation projects and grants: 3.2054.2014/K, 11.G34.31.0029 and 02.B.49.21.0003. The research was supported by RFBR Grant 14-02-31809 and 14-02-00625.
\end{acknowledgments}

\bibliography{Neutrons}

\end{document}